\documentstyle[aps,preprint,tighten]{revtex}

\begin{document}


\draft

\title{Wave function of a Brownian particle}
\author{R. M. Cavalcanti\footnote{Electronic address: rmc@itp.ucsb.edu}}
\address{Institute for Theoretical Physics, University of California, 
Santa Barbara, CA 93106-4030}
\maketitle

\begin{abstract}

Using the Hamiltonian of Caldirola [Nuovo Cimento {\bf 18}, 393 (1941)]
and Kanai [Prog.\ Theor.\ Phys.\ {\bf 3}, 440 (1948)], 
we study the time evolution of the wave function of a particle
whose classical motion is governed by the Langevin equation
$m\ddot{x}+\eta\dot{x}=F(t)$. We show in particular
that if the initial wave function is Gaussian, then (i) it remains Gaussian
for all times, (ii) its width grows, approaching a finite value
when $t\to\infty$, and (iii) its center describes a Brownian motion
and so the uncertainty in the position of the particle grows without
limit.
 
\end{abstract}





One of the starting points for the theory of the
Brownian motion is the Langevin equation
\begin{equation}
\label{Langevin}
m\ddot{x}+\eta\dot{x}+\frac{\partial V}{\partial x}=F(t),
\end{equation}
where $m$ is the mass of the particle, $\eta$
is a damping constant, $V(x)$ is the potential acting
on the particle and $F(t)$ is a Gaussian random
force, obeying the relations
\begin{equation}
\overline{F(t)}=0, \qquad
\overline{F(t)F(t')}=\phi(t-t'),
\end{equation}
where $\phi(t)$ is a function sharply peaked at $t=0$
and the overbar represents the average over noise.

The need to quantize such a system appears naturally
when one studies, for instance, quantum electrodynamics in cavities,
the low-temperature behavior of Josephson junctions
or the effects of dissipation on quantum tunneling
(for references, see\cite{Caldeira,Weiss}). 
In the system-plus-reservoir approach\cite{Caldeira,Weiss},
one treats the system in which one is interested
and its environment, the ``reservoir,'' as a closed composite 
system, applies the usual rules of quantization, and
then eliminates the reservoir degrees of freedom.

A more phenomenological approach consists in the
quantization of the Caldirola-Kanai (CK) time-dependent
Hamiltonian\cite{Caldirola},
\begin{equation}
H=e^{-\gamma t}\,\frac{p^2}{2m}+e^{\gamma t}\,[V(x)-F(t)x]
\qquad\left(\gamma\equiv\frac{\eta}{m}\right),
\end{equation}
from which Eq.~(\ref{Langevin}) results as a consequence
of the Hamilton equations.
Making the usual association $p=-i\partial/\partial x$
(in units such that $\hbar=1$),
one then obtains the Schr\"odinger equation 
\begin{equation}
\label{Sch1}
i\,\frac{\partial}{\partial t}\,\psi(x,t)=\left\{
-\frac{e^{-\gamma t}}{2m}\,\frac{\partial^2}{\partial x^2}
+e^{\gamma t}\,[V(x)-F(t)x]\right\}\psi(x,t).
\end{equation}

This approach has been severely criticized by some 
authors\cite{Menon}, but Caldirola and Lugiato\cite{Caldirola2}
have shown that, in the case of the damped harmonic oscillator
with a stochastic force, it gives the the same results as
the more orthodox approach based on a master equation
derived by elimination of the degrees of freedom of the thermal
reservoir. Their analysis can be generalized for an
arbitrary potential $V(x)$, if one uses the Caldeira-Leggett
model\cite{Caldeira} to describe the interaction of the particle
with its environment. One can then derive a quantum-mechanical Langevin 
equation for the position operator of the particle with an
operator-valued stochastic force\cite{Yu}. The CK Hamiltonian
can then be viewed as an effective one-particle Hamiltonian
that, through the Heisenberg equations of motion
for $x$ and $p$, generates the Langevin equation for $x$.
One should note, however, that because $F(t)$ is an operator
in the Hilbert space of the environment,
the wave function $\psi(x,t)$ is also an operator in that
space since it is a functional of $F(t)$. Therefore,
in order to obtain expectation values of operators defined
in the Hilbert space of the particle, 
one must know not only the wave function $\psi(x,t)$\cite{Landau},
but also the state of the reservoir, as specified by its
density operator, which in turn determines the
noise correlation function $\phi(t)$. [In practice, 
if one is not interested in short-time effects,
one can assume a white noise, i.e., $\phi(t)=D\,\delta(t)$,
where $D$ is a temperature-dependent coefficient
determined in accordance with the fluctuation-dissipation theorem.]
Interpreted this way, the Caldirola-Kanai approach
to the quantum Brownian motion is akin to the quantum
state diffusion picture proposed by Gisin and Percival\cite{Gisin},
in which a master equation for the density operator $\rho$ is
replaced with a stochastic equation of motion for the
state vector $|\psi\rangle$; the former is then recovered
by averaging $|\psi\rangle\langle\psi|$ over the fluctuations.
In what follows, we shall illustrate these ideas by
studying the evolution of a wave packet in the case $V=0$.



In order to solve Eq.~(\ref{Sch1}) (with $V=0$), 
we first make the change of variable
\begin{equation}
\label{tau}
\tau(t)=\frac{1}{\gamma}\left(1-e^{-\gamma t}\right).
\end{equation}
Equation (\ref{Sch1}) then becomes 
\begin{equation}
\label{Sch2}
i\,\frac{\partial}{\partial\tau}\,\psi(x,\tau)=\left[
-\frac{1}{2m}\,\frac{\partial^2}{\partial x^2}-f(\tau)x\right]\psi(x,\tau),
\end{equation}
where
\begin{equation}
f(\tau)=e^{2\gamma t}F(t)=\frac{F\left(-\gamma^{-1}
\ln(1-\gamma\tau)\right)}{(1-\gamma\tau)^2}.
\end{equation}
Equation (\ref{Sch2}) is the Schr\"odinger equation of
a particle in a time-dependent
electric field, for which solutions can be found the form of plane 
waves\cite{Fallieros}. In fact, 
\begin{equation}
\psi(x,\tau)=N\,e^{ip(\tau)x-i\alpha(\tau)}
\end{equation}
is a solution of Eq.\ (\ref{Sch2}) provided $p(\tau)$ and
$\alpha(\tau)$ satisfy 
\begin{equation}
\frac{dp}{d\tau}=f(\tau),\qquad
\frac{d\alpha}{d\tau}=\frac{p^2}{2m}.
\end{equation}
The solutions of these equations (and the corresponding
wave functions) can be labeled by the initial value of
$p(\tau)$, which we denote by $k$,
\begin{mathletters}
\begin{eqnarray}
p_k(\tau)&=&k+\int_0^{\tau}f(\tau')\,d\tau'\equiv k+I(\tau),
\\
\alpha_k(\tau)&=&\frac{1}{2m}\left[k^2\tau+2k\int_0^{\tau}
I(\tau')\,d\tau'+\int_0^{\tau}I^2(\tau')\,d\tau'\right]
\nonumber \\
&\equiv&\frac{k^2\tau}{2m}+kf_1(\tau)+f_2(\tau).
\end{eqnarray}
\end{mathletters} 
[The initial value of $\alpha(\tau)$ can
be absorbed in the normalization constant $N$
and so without lack of generality, we may take $\alpha(0)=0$.]
With $N=(2\pi)^{-1/2}$, one can easily verify that
\begin{mathletters}
\begin{eqnarray}
\int_{-\infty}^{\infty}\psi_k^*(x,\tau)\,\psi_{k'}(x,\tau)\,dx=\delta(k-k'),
\\
\int_{-\infty}^{\infty}\psi_k(x,\tau)\,\psi_k^*(x',\tau)\,dk=\delta(x-x'). 
\end{eqnarray}
\end{mathletters}
The second identity allows one to write the propagator as
\begin{equation}
\label{G}
G(x,t|x',t')=\int_{-\infty}^{\infty}\psi_k(x,\tau)\,\psi_k^*(x',\tau')\,dk,
\end{equation}
where $\tau=\tau(t)$, $\tau'=\tau(t')$.

Using the above result, let us investigate the time evolution
of the Gaussian wave packet
\begin{equation}
\psi(x,0)=(2\pi\sigma^2)^{-1/4}e^{-x^2/4\sigma^2}.
\end{equation}
Performing the integration over $k$ in Eq.\ (\ref{G}), one finds 
\begin{equation}
G(x,t|x',0)=\sqrt{\frac{m}{2\pi i\tau}}\,
\exp\left\{\frac{im[x-x'-f_1(\tau)]^2}
{2\tau}+iI(\tau)x-if_2(\tau)\right\},
\end{equation} 
so that
\begin{eqnarray}
\psi(x,t)&=&\int_{-\infty}^{\infty}G(x,t|x',0)\,\psi(x',0)\,dx'
\nonumber \\
&=&(2\pi)^{-1/4}\left(\sigma+\frac{i\tau}{2m\sigma}\right)^{-1/2}
\exp\left\{-\frac{m[x-f_1(\tau)]^2}
{4m\sigma^2+2i\tau}+iI(\tau)x-if_2(\tau)\right\}.
\end{eqnarray}
The probability density is then given by
\begin{equation}
\label{P}
|\psi(x,t)|^2=\frac{1}{\sqrt{2\pi}\sigma(t)}\,
\exp\left\{-\frac{[x-f_1(\tau)]^2}{2\sigma^2(t)}\right\},
\end{equation}
where 
\begin{equation}
\sigma(t)=\sqrt{\sigma^2+\frac{\tau^2(t)}{4m^2\sigma^2}}.
\end{equation}
The spreading of the wave packet is the same one
found in\cite{Yu,Hasse} taking into account only the dissipation. 

At first sight this result may seem
paradoxical: Even in the presence of a fluctuating force, the
uncertainty in position  
\begin{equation}
\label{Dxqu}
(\Delta x)_{qu}\equiv\sqrt{\langle x^2\rangle-\langle x\rangle^2}=\sigma(t)
\end{equation}
tends to a finite value when $t\to\infty$. However, it should be noted that
the above quantity measures the uncertainty in the position of the particle
with respect to the ``center of mass'' of the wave packet, 
$\langle x(t)\rangle$, which cannot be determined precisely:
According to Ehrenfest theorem, 
$\langle x(t)\rangle$ satisfies Eq.~(\ref{Langevin}) and so
describes a Brownian motion. Therefore,
there is a ``classical'' uncertainty in the position of the center
of the wave packet that, in the case of a white noise, 
is given by\cite{Reif}
\begin{equation}
\label{Dxcl}
(\Delta x)_{cl}\equiv\sqrt{\overline{\langle x\rangle^2}
-\overline{\langle x\rangle}^2}=\sqrt{\frac{D}{\eta^2}\,
[t-\gamma^{-1}(1-e^{-\gamma t})]},
\end{equation}
which, in contrast with the ``quantum'' uncertainty in the
position of the particle, does not ``freeze'' when $t\to\infty$.

The above definitions of quantum and classical uncertainties
are somewhat artificial since they cannot be directly compared with
experiment. (This would require not only an ensemble of 
particles with the same initial state, which one could possibly
manage to prepare, but also an ensemble of reservoirs 
in the same microstate.) As argued before, the natural
definition of the expectation value of an operator ${\cal O}$ 
(i.e., the one that can be compared with experiment) is given by 
$\overline{\langle{\cal O}\rangle}$ and so the 
uncertainty in the position of the particle is given by
\begin{equation}
\label{Dx}
\Delta x\equiv\sqrt{\overline{\langle x^2\rangle}-
\overline{\langle x\rangle}^2}
=\sqrt{(\Delta x)^2_{qu}+(\Delta x)^2_{cl}}.
\end{equation}

In conclusion, I would like to emphasize that the whole point of
the above exercise is to show that, {\em provided fluctuation effects
are properly taken into account}, the CK approach to
the quantum Brownian motion can give the same results as
the more conventional approaches based on master equations or 
influence functionals.  


\bigskip

I thank L.~Davidovich and C.~A.~A.~de Carvalho
for a critical reading of this paper. 
This work was supported by the Conselho Nacional de
Desenvolvimento Cient\'\i fico e Tecnol\'ogico (CNPq) and, in part, by the
National Science Foundation under Grant No.~PHY94-07194.  




\begin{references}

\bibitem{Caldeira} A. O. Caldeira and A. J. Leggett, Physica A {\bf 121},
587 (1983).

\bibitem{Weiss} U. Weiss, {\it Quantum Dissipative Systems\/} 
(World Scientific, Singapore, 1993).

\bibitem{Caldirola} P. Caldirola, Nuovo Cimento {\bf 18}, 393 (1941);
E. Kanai, Prog. Theor. Phys. {\bf 3}, 440 (1948).

\bibitem{Menon} Most of these criticisms are rebutted (properly, in my
opinion) by V. J. Menon, N. Chanana and Y. Singh, Prog. Theor. Phys. 
{\bf 98}, 321 (1997).

\bibitem{Caldirola2} P. Caldirola and L. A. Lugiato, Physica A {\bf 116},
248 (1982).

\bibitem{Yu} C. P. Sun and L. H. Yu, Phys. Rev. A {\bf 51}, 1845 (1995).

\bibitem{Landau} This is another way to say that a Brownian particle 
(or, more generally, an open system) does not have a wave function.
See L. D. Landau and E. M. Lifshitz, {\it Quantum Mechanics\/},
3rd. ed. (Pergamon, Oxford, 1977), \S14.

\bibitem{Gisin} N. Gisin and I. C. Percival, J. Phys. A {\bf 25}, 5677 (1992).

\bibitem{Fallieros} S. Fallieros and J. L. Friar,
Am. J. Phys. {\bf 50}, 1001 (1992).

\bibitem{Hasse} R. W. Hasse, J. Math. Phys. {\bf 16}, 2005 (1975).

\bibitem{Reif} F. Reif, {\it Fundamentals of Statistical and Thermal
Physics\/} (McGraw-Hill, New York, 1965), Chap. 15. [It should be
emphasized that this result is derived under the assumption of a
white-noise. For instance, in Sec.~7 of Ref.\cite{Caldeira} the authors
show that when $T\to 0$, $(\Delta x)^2\sim\ln t$ when $t\to\infty$,
in contrast with the behavior predicted by Eq.~(\ref{Dxcl}). 
This discrepancy occurs because in their model for
the reservoir the noise becomes ``colored'' when $T\to 0$.]

\end{references}
\end{document}